\documentclass[a4paper]{jpconf}

\usepackage{amsmath,amssymb}
\usepackage[dvips]{graphicx}
\usepackage{cite}

\newcommand{\Dmq}{\ensuremath{\Delta m^2}}
\newcommand{\Sol}{\textsc{sol}}
\newcommand{\Atm}{\textsc{atm}}
\newcommand{\Sbl}{\textsc{lsnd}}

\begin{document}

\title{Sterile neutrinos after the first MiniBooNE results}

\author{Michele Maltoni}
\address{%
  Departamento de F\'isica Te\'orica \& Instituto de F\'isica
  Te\'orica UAM/CSIC, Facultad de Ciencias C-XI, Universidad
  Aut\'onoma de Madrid, Cantoblanco, E-28049 Madrid, Spain}
\ead{michele.maltoni@uam.es}

\begin{abstract}
    In view of the recent results from the MiniBooNE experiment we
    revisit the global neutrino oscillation fit to short-baseline
    neutrino data by adding one, two or three sterile neutrinos with
    eV-scale masses to the three Standard Model
    neutrinos~\cite{Maltoni:2007zf}.
    We find that four-neutrino oscillations of the (3+1) type, which
    have been only marginally allowed before the recent MiniBooNE
    results, become even more disfavored with the new data.
    In the framework of so-called (3+2) five-neutrino mass schemes the
    MiniBooNE results can be nicely reconciled with the LSND
    appearance evidence thanks to the possibility of CP violation
    available in such oscillation schemes; however, the tension
    between appearance and disappearance experiments represents a
    serious problem in (3+2) schemes, so that these models are
    ultimately not viable.
    This tension remains also when a third sterile neutrino is added,
    and we do not find a significant improvement of the global fit in
    a (3+3) scheme.
\end{abstract}

\section{Introduction}

Recently the first results from the MiniBooNE (MB)
experiment~\cite{MB-talk, AguilarArevalo:2007it} at Fermilab have been
released on a search for $\nu_\mu\to\nu_e$ appearance with a baseline
of 540~m and a mean neutrino energy of about 700~MeV. The primary
purpose of this experiment is to test the evidence of $\bar\nu_\mu \to
\bar\nu_e$ transitions reported by the LSND experiment at Los
Alamos~\cite{Aguilar:2001ty} with a very similar $L/E$ range.
Reconciling the LSND signal with the other evidence for neutrino
oscillations is a long-standing challenge for neutrino phenomenology,
since the mass-squared differences required to explain the solar,
atmospheric and LSND experimental results in terms of neutrino
oscillations differ from one another by various orders of magnitude.
Consequently, there is no consistent way to explain all these three
signals invoking only oscillations among the three known neutrinos.
Therefore, in order to explain the LSND anomaly one had to invoke an
extension of the three-neutrino mixing scenario, introducing either a
mechanism to generate at least a third mass-square difference, or a
new form of flavor transition beyond oscillations. Following
Ref.~\cite{Maltoni:2007zf}, in this talk we will concentrate on the
first possibility, starting from models with one extra sterile
neutrino (Sec.~\ref{sec:fourmix}) and then considering models with two
and three sterile neutrino states (Sec.~\ref{sec:moremix}).

\section{Four-neutrino mixing}
\label{sec:fourmix}

In four-neutrino models, one extra sterile state is added to the three
weakly interacting ones. The relation between the flavor and the mass
eigenstates can be described in terms of a $4 \times 4$ unitary matrix
$U$, which generalizes the usual $3 \times 3$ leptonic matrix of the
Standard Model. There are six possible four-neutrino schemes that can
accommodate the results from solar and atmospheric neutrino
experiments and contain a third much larger $\Dmq$.  They can be
divided into two classes: (3+1) and (2+2). In the (3+1) schemes, there
is a group of three close-by neutrino masses that is separated from
the fourth one by the larger gap.  In (2+2) schemes, there are two
pairs of close masses separated by the large gap. While different
schemes within the same class are presently indistinguishable, schemes
belonging to different classes lead to very different phenomenological
scenarios.

\bigskip

A characteristic feature of (2+2) schemes is that the extra sterile
state cannot be simultaneously decoupled from \emph{both} solar and
atmospheric oscillations. To understand why, let us define $\eta_s =
\sum_{i \,\in\, \Sol} |U_{s i}|^2$ and $c_s = \sum_{j \,\in\, \Atm}
|U_{s j}|^2$, where the sums in $i$ and $j$ run over mass eigenstates
involved in solar and atmospheric neutrino oscillations, respectively.
Clearly, the quantities $\eta_s$ and $c_s$ describe the fraction of
sterile neutrino relevant for each class of experiment. 
Results from atmospheric and solar neutrino data imply that in both
kind of experiments oscillation takes place mainly between active
neutrinos. Specifically, from Fig.~46 of
Ref.~\cite{GonzalezGarcia:2007ib} we get $\eta_s \le 0.31$ and $c_s
\le 0.36$ at the $3\sigma$ level. However, in (2+2) schemes unitarity
implies $\eta_s + c_s = 1$.
A statistical analysis using the \emph{parameter goodness of fit} (PG)
proposed in~\cite{Maltoni:2003cu} gives $\chi^2_\text{PG} = 30.7$ for
1 d.o.f., corresponding to a $5.5\sigma$ rejection ($\text{PG} =
3\times 10^{-8}$) of the (2+2) hypothesis.
These models are therefore ruled out at a very high confidence level,
and in the rest of this talk we will not consider them anymore.

\bigskip

\begin{figure}\centering
    \includegraphics[width=0.95\textwidth]{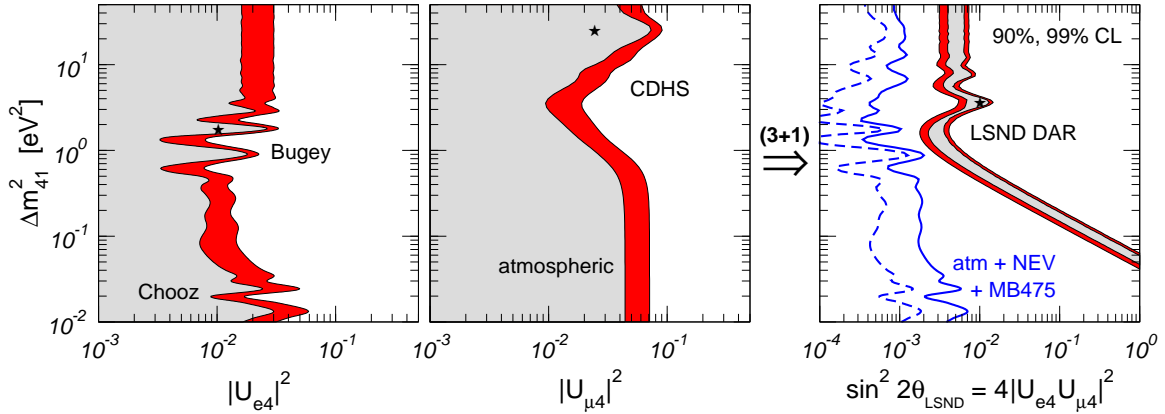}
    \caption{\label{fig:disapp}%
      Bounds on $|U_{e4}|^2$ (left panel), on $|U_{\mu 4}|^2$ (middle
      panel) and on $\sin^2 2\theta$ (right panel) in (3+1) schemes,
      as a function of $\Dmq_{41}$. Different contours correspond to
      90\% and 99\% CL.}
\end{figure}

On the other hand, (3+1) schemes are not affected by this problem.
Although the experimental bounds on $\eta_s$ and $c_s$ quoted above
still hold, the condition $\eta_s + c_s = 1$ no longer applies. For
what concerns neutrino oscillations, in (3+1) models the mixing
between the sterile neutrino and the three active ones can be reduced
at will, and in particular it is possible to recover the usual
three-neutrino scenario as a limiting case. However, as widely
discussed in the literature (see, \textit{e.g.},
Ref.~\cite{Maltoni:2002xd} and references therein) these models are
strongly disfavored as an explanation of LSND by the data from other
short-baseline (SBL) laboratory experiments.
In the limit $\Dmq_\Sbl \gg \Dmq_\Atm \gg \Dmq_\Sol$ the probability
$P_{\nu_\mu\to\nu_e}$ which is relevant for LSND as well as for
KARMEN~\cite{Armbruster:2002mp}, NOMAD~\cite{Astier:2003gs} and
MiniBooNE is driven by the large $\Dmq_{41}$, and is given by 
\begin{equation}
    \label{eq:Pmue}
    P_{\nu_\mu \to \nu_e} = P_{\bar\nu_\mu \to \bar\nu_e} =
    4 \, |U_{e4} U_{\mu 4}|^2 \, \sin^2 \frac{\Dmq_{41} L}{4E} \,,
\end{equation}
where $L$ is the distance between source and detector.  The LSND,
KARMEN, NOMAD and MiniBooNE experiments give allowed regions in the
$(\Dmq_{41},\, |U_{e4} U_{\mu4}|^2)$ plane which can be directly
obtained from the corresponding two-neutrino exclusion
plots~\cite{Aguilar:2001ty, Armbruster:2002mp, Astier:2003gs,
AguilarArevalo:2007it}. At the light of the recent MiniBooNE result
which is consistent with no oscillations above 475~MeV, practically
all the LSND region is now excluded.
In addition, further constraints on $|U_{e4} U_{\mu4}|^2$ can be
obtained by combining together the bounds on $|U_{e4}|$ and
$|U_{\mu4}|$ derived from reactor and accelerator experiments (mainly
Bugey~\cite{Declais:1994su} and CDHS~\cite{Dydak:1983zq}) as well as
solar and atmospheric data. Present bounds are plotted in the two
leftmost panels of Fig.~\ref{fig:disapp} as a function of $\Dmq_{41}$;
for a discussion on the sensitivity of future experiments, see
Ref.~\cite{Donini:2007yf}.
The results of the global analysis presented in
Ref.~\cite{Maltoni:2007zf}, which includes atmospheric and
long-baseline data together with all the short-baseline experiments
observing \emph{no evidence} (NEV), are summarized in the right panel
of Fig.~\ref{fig:disapp}.
Using the PG test discussed above we find $\chi^2_\text{PG} = 24.7$
for 2 d.o.f., corresponding to a $4.6\sigma$ rejection ($\text{PG} =
4\times 10^{-6}$) of the (3+1) hypothesis. These results show that
(3+1) schemes are now ruled out as a possible explanation of
LSND~\cite{Maltoni:2007zf}.
In addition, it should be noted that the low-energy excess observed by
MiniBooNE at $E_\nu \le 475$~MeV cannot be explained in terms of
oscillations with only one large mass-squared difference, thus adding
another problem to these models in case this excess is confirmed to be
a real signal.

\section{Five-neutrino and six-neutrino mixing}
\label{sec:moremix}

Five-neutrino schemes of the (3+2) type are a straight-forward
extension of (3+1) schemes. In addition to the cluster of the three
neutrino mass states accounting for ``solar'' and ``atmospheric'' mass
splittings now two states at the eV scale are added, with a small
admixture of $\nu_e$ and $\nu_\mu$ to account for the LSND signal.
In the Appendix of Ref.~\cite{Peres:2000ic} it was suggested that such
models could somewhat relax the tension existing between
short-baseline experiments and the LSND data. In
Ref.~\cite{Sorel:2003hf} a complete analysis was performed, finding
that indeed the disagreement between LSND and null-result experiments
is reduced. Here we will reconsider this possibility at the light of
the new MiniBooNE data.
As explained in Ref.~\cite{AguilarArevalo:2007it}, MiniBooNE found no
evidence of oscillations above 475~MeV, whereas below this energy a
$3.6\sigma$ excess of $96 \pm 17 \pm 20$ events is observed. Whether
this excess comes indeed from $\nu_\mu\to\nu_e$ transitions or has
some other origin is under investigation~\cite{AguilarArevalo:2007it}.
Lacking any explanation in terms of backgrounds or systematical
uncertainties, we will present the results obtained using both the
full energy range from 300~MeV to 3~GeV (``MB300'') and for the
restricted range from 475~MeV to 3~GeV (``MB475'').

\bigskip

\begin{figure}\centering 
    \includegraphics[width=0.9\textwidth]{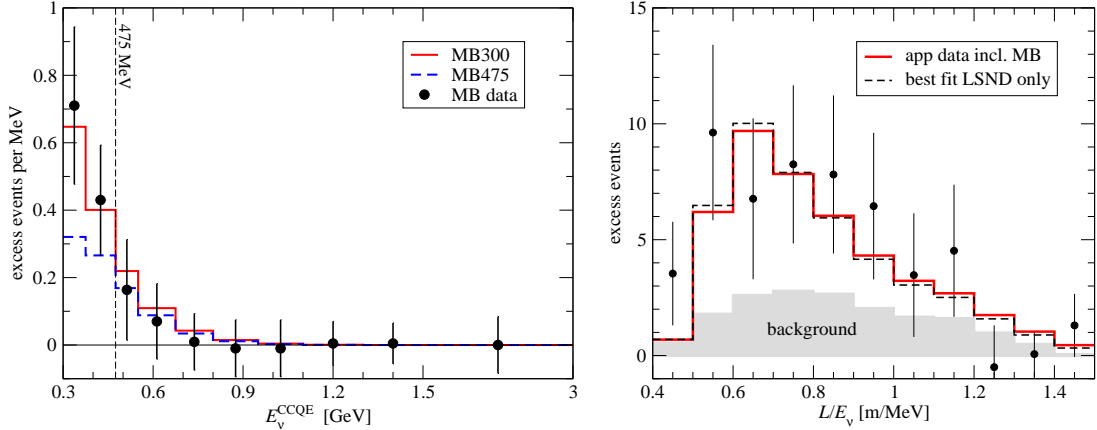}
    \caption{\label{fig:spectrum}%
      Spectral data for the MiniBooNE (left) and LSND (right)
      experiments, calculated at the best fit point of a combined fit
      of LSND, KARMEN, NOMAD and MB in (3+2) schemes.}
\end{figure}

As for (3+1) models, in (3+2) schemes the \emph{appearance} data
(LSND, KARMEN, NOMAD, and MiniBooNE) can be described using the SBL
approximation $\Dmq_\Sol \approx 0$ and $\Dmq_\Atm \approx 0$, in
which case the relevant transition probability is given by
\begin{multline}
    \label{eq:5nu-prob}
    P_{\nu_\mu\to\nu_e} =
    4 \, |U_{e4} U_{\mu 4}|^2 \, \sin^2 \phi_{41} +
    4 \, |U_{e5} U_{\mu 5}|^2 \, \sin^2 \phi_{51}
    \\
    + 8 \,|U_{e4} U_{\mu 4}| \, |U_{e5} U_{\mu 5}| \,
    \sin\phi_{41}\sin\phi_{51}\cos(\phi_{54} - \delta) \,,
\end{multline}
with the definitions $\phi_{ij} \equiv \Dmq_{ij} L / 4E$ and $\delta
\equiv \arg( U_{e4}^* U_{\mu 4} U_{e5} U_{\mu 5}^* )$.
Eq.~\eqref{eq:5nu-prob} holds for neutrinos (NOMAD and MB); for
anti-neutrinos (LSND and KARMEN) one has to replace $\delta \to
-\delta$. Note that Eq.~\eqref{eq:5nu-prob} is invariant under the
transformation $4\leftrightarrow 5$ and $\delta \leftrightarrow
-\delta$, and depends only on the combinations $|U_{e4}U_{\mu 4}|$ and
$|U_{e5}U_{\mu 5}|$.
An important observation is that non-trivial values of the complex 
phase $\delta$ lead to CP violation, and hence in (3+2) schemes much
more flexibility is available to accommodate the results of LSND
(anti-neutrinos) and MB (neutrinos). In Fig.~\ref{fig:spectrum} we
show the prediction for MB and LSND at the best fit points of the
combined MB, LSND, KARMEN, NOMAD analysis. As can be seen from this
figure, MB data can be fitted very well while simultaneously
explaining the LSND evidence. Furthermore, in this case also the low
energy MB data can be explained, and therefore, in contrast to (3+1)
schemes, (3+2) oscillations offer an appealing possibility to account
for this excess. For both MB475 and MB300 a goodness-of-fit of 85\% is
obtained, showing that MB is in very good agreement with global SBL
appearance data including LSND.

\bigskip

\begin{figure}\centering 
    \includegraphics[width=0.85\textwidth]{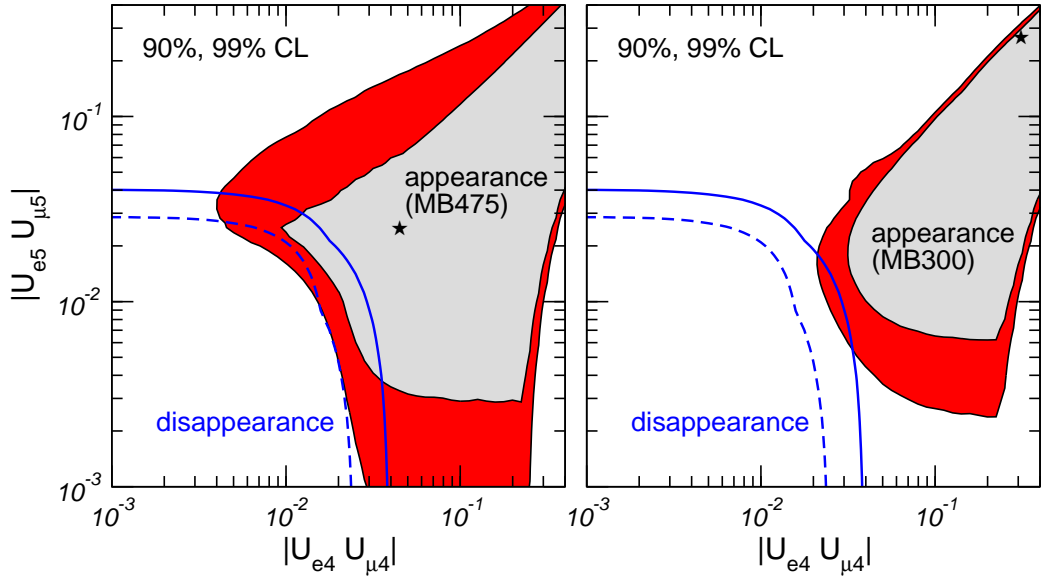}
    \caption{\label{fig:regions}%
      Allowed regions at 90\% and 99\%~CL in (3+2) schemes from the
      analysis of appearance and disappearance data, projected onto
      the plane of $|U_{e4} U_{\mu 4}|$ and $|U_{e5} U_{\mu 5}|$.}
\end{figure}

On the other hand, once \emph{disappearance} data are included in the
analysis, the quality of the fit decreases considerably. Indeed, even
in (3+2) schemes short-baseline experiments pose stringent bounds on
the mixing angles $|U_{ei}|$ and $|U_{\mu i}|$, in close analogy with
(3+1) models described in Sec.~\ref{sec:fourmix} and shown in the two
leftmost panels of Fig.~\ref{fig:disapp}. Since rather large values of
$|U_{e4} U_{\mu 4}|$ and $|U_{e5} U_{\mu 5}|$ are needed to account
for the negative result of MiniBooNE as well as the positive signal of
LSND, one expects that reconciling appearance and disappearance data
will be a problem also within (3+2) models.
This tension is illustrated in Fig.~\ref{fig:regions}, where the
projections of the allowed regions in the plane of the appearance
amplitudes $|U_{e4} U_{\mu 4}|$ and $|U_{e5} U_{\mu 5}|$ are shown. 
Indeed the opposite trend of the two data sets is clearly visible,
especially when the low energy excess in MB is included (right panel).
In order to quantify this disagreement one can apply the PG test to
appearance versus disappearance data without MB, with MB475, and with
MB300:
\begin{equation}\label{eq:5nuPG}
    \text{APP vs DIS:} \quad \left\lbrace
    \begin{aligned}
	\chi^2_\text{PG} &= 17.5\,, \quad
	&\text{PG} &= 1.5\times 10^{-3} \qquad \text{(no MB),}
	\\
	\chi^2_\text{PG} &= 17.2 \,, \quad 
	& \text{PG} &= 1.8\times 10^{-3} \qquad \text{(MB475),}
	\\
	\chi^2_\text{PG} &= 25.1 \,, \quad
	& \text{PG} &= 4.8\times 10^{-5} \qquad \text{(MB300).}
    \end{aligned} \right.
\end{equation}
From these numbers we conclude that also in (3+2) schemes the tension
between appearance and disappearance experiments is quite severe. If
MB475 is used the result is very similar to the pre-MiniBooNE
situation implying inconsistency at about $3.1\sigma$, whereas in case
of the full MB300 data the tension becomes significantly worse (about
$4\sigma$), since appearance data are more constraining because of the
need to accommodate LSND as well as the MB excess at low energies.

\bigskip

Finally, since there are three active neutrinos it seems natural to
consider also the case of three sterile neutrinos. If all three
additional neutrino states have masses in the eV range and mixings as
relevant for the SBL experiments under consideration, such a model
will certainly have severe difficulties to accommodate standard
cosmology~\cite{Hannestad:2006mi}.
Besides this fact, the results of the search performed in 
Ref.~\cite{Maltoni:2007zf} show that there is only a marginal
improvement of the fit by 1.7 units in $\chi^2$ for MB475 (3.5 units
for MB300) with respect to (3+2), to be compared with four additional
parameters in the model. Hence, the conclusion is that here are no
qualitatively new effects in the (3+3) scheme. The conflict between
appearance and disappearance data remains a problem, and the
additional freedom introduced by the new parameters does not relax
significantly this tension.

\section{Conclusions}

We have considered the global fit to short-baseline neutrino
oscillation data including the recent data from MiniBooNE, in the
framework of (3+1), (3+2) and (3+3) oscillation models. Four-neutrino
models are ruled out since (a) the don't allow to account for the low
energy event excess in MB, (b) MiniBooNE result cannot be reconciled
with LSND, and (c) there is severe tension between \emph{appearance}
and \emph{disappearance} experiments. Five-neutrino models provide a
nice way out for problems (a) and (b), but fail to resolve (c).
Similarly, six-neutrino models do not offer qualitatively new effects
with respect to (3+2). In all cases we find severe tension between
different sub-samples of the data, hence we conclude that at the light
of present experimental results it is \emph{not} possible to explain
the LSND evidence in terms of sterile neutrinos.

\section*{Acknowledgments}

Work supported by MCYT through the Ram\'on y Cajal program, by
CiCYT through the project FPA-2006-01105 and by the Comunidad
Aut\'onoma de Madrid through the project P-ESP-00346.

\end{document}